\documentclass[twocolumn,showpacs,preprintnumbers,amsmath,amssymb,floatfix,a4paper]{revtex4}

\usepackage{graphicx}
\usepackage{dcolumn}
\usepackage{bm}
\usepackage{setspace}
\usepackage{amsmath}

\begin{document}

\title{Non-Extensive Statistics, New Solution to the Cosmological Lithium Problem}

\author{J.J.~He$^1$}
\email{jianjunhe@impcas.ac.cn}
\author{S.Q.~Hou$^{1,2}$}
\author{A.~Parikh$^{3,4}$}
\author{D.~Kahl$^{5}$}
\author{C.A.~Bertulani$^{6}$}
\author{other collaborators}

\affiliation{$^1$Key Laboratory of High Precision Nuclear Spectroscopy and Center for Nuclear Matter Science, Institute of Modern Physics, Chinese Academy of Sciences, Lanzhou 730000, China}
\affiliation{$^2$University of Chinese Academy of Sciences, Beijing 100049, China}
\affiliation{$^3$Departament de F\'{\i}sica i Enginyeria Nuclear, EUETIB, Universitat Polit\`{e}cnica de Catalunya, Barcelona E-08036, Spain}
\affiliation{$^4$Institut d'Estudis Espacials de Catalunya, Barcelona E-08034, Spain}
\affiliation{$^5$Center for Nuclear Study, The University of Tokyo, RIKEN campus, Wako, Saitama 351-0198, Japan}
\affiliation{$^6$Texas A\&M University-Commerce, Commerce, TX 75429-3011, USA}

\date{\today}

\begin{abstract}
In the primordial Big Bang nucleosynthesis (BBN), only the lightest nuclides (D, $^3$He, $^4$He, and $^7$Li) were synthesized in appreciable quantities, and
these relics provide us a unique window on the early universe. Currently, BBN simulations give acceptable agreement between theoretical and observed abundances
of D and $^4$He, but it is still difficult to reconcile the predicted $^7$Li abundance with the observation for the Galactic halo stars. The BBN model
overestimates the primordial $^7$Li abundance by about a factor of three, so called the cosmological lithium problem, a long-lasting pending issue in BBN.
Great efforts have been paid in the past decades, however, the conventional nuclear physics seems unable to resolve such problem. It is well-known that
the classical Maxwell-Boltzmann (MB) velocity distribution has been usually assumed for nuclei in the Big-Bang plasma. In this work, we have thoroughly
investigated the impact of non-extensive Tsallis statistics (deviating from the MB) on thermonuclear reaction rates involved in standard models of BBN.
It shows that the predicted primordial abundances of D, $^4$He, and $^7$Li agree very well with those observed ones by introducing a non-extensive parameter
$q$. It is discovered that the velocities of nuclei in a hot Big-Bang plasma indeed violate the classical Maxwell-Boltzmann (MB) distribution in
a very small deviation of about 6.3--8.2\%. Thus, we have for the first time found a new solution to the cosmological lithium problem without introducing any
mysterious theories. Furthermore, the implications of non-extensive statistics in other exotic high-temperature and density astrophysical environments should
be explored, which might offer new insight into the nucleosynthesis of heavy elements.
\end{abstract}


\maketitle

The primordial Big-Bang Nucleosynthesis (BBN) began when the universe was 3-minutes old and ended less than half an hour later when the nuclear reactions were
quenched by the low temperature and density conditions in the expanding universe. Only the lightest nuclides (D, $^3$He, $^4$He, and $^7$Li) were synthesized
in appreciable quantities through BBN, and these relics provide us a unique window on the early universe. Currently, standard BBN simulations give acceptable
agreement between theoretical and observed abundances of D and $^4$He, but it is still difficult to reconcile the predicted $^7$Li abundance with the
observation for the Galactic halo stars (GHS). The BBN model overestimates observations interpreted as primordial $^7$Li abundance by about a factor of
three~\cite{Li2,Li}, so called the cosmological lithium problem, a pending issue in BBN for a very long time. This apparent discrepancy has promoted a wealth of
experimental and theoretical inquiries. However, conventional nuclear physics seems unable to resolve such problem (e.g., see
Refs.~\cite{Cyburt08,boyd,kirsebom,hamm,pizz}). Although this problem seems to be solved by introducing a long-live massive $X$ particles
recently~\cite{kusa13}, these particles and the relevant theories are too mysterious.
Furthermore, another model~\cite{coc13} in which neutrons can oscillate into mirror neutrons has been recently proposed. Such a mechanism allows for an effective
late time neutron injection, which induces an increase of the destruction of $^7$Be, due to an increase of the neutron capture, and then a decrease of the final
$^7$Li abundance. However, this theory still cannot reconcile all the primordial abundances whatever the value of the oscillation time.

In the BBN model, the predominant nuclear-physics inputs are thermonuclear reaction rates (derived from cross sections). In the past decades, great efforts have
been undertaken to determine these data with high accuracy (e.g., see compilations~\cite{fcz67,Wagoner67,wagoner69,cf88,Smiths93,Angulo99,Des04,Serpico04,yixu}).
A key assumption in all thermonuclear rate determinations is that the velocities of ions may be described by the classical Maxwell-Boltzmann (MB)
distribution~\cite{mandl08,Rolfs88}. It is well-known that the MB distribution was derived for describing the thermodynamic equilibrium properties of the ideal
gas, where the particles move freely without interacting with one another, except for very brief collisions in which they exchange energy and momentum with each
other or with their thermal environment. This classical distribution was ultimately verified by a high-resolution experiment~\cite{miller55} at temperatures
around 900 K. However, in real gases, there are various effects (e.g., van der Waals interactions, relativistic speed limits, etc.) that make their speed
distribution sometimes very different from the MB form. Moreover, stellar systems are generally subject to spatial long-range interactions, causing the
thermodynamics of many-body self-gravitating systems to show some peculiar features differing drastically from typical ones~\cite{tar02}. Therefore it is worth
asking: does the nucleus still obey the classical MB distribution in the extremely complicated Big-Bang plasma environments?

To address such issues, one may utilize the Tsallis statistics (also referred to as non-extensive statistics)~\cite{Tsallis88,Tsal09}, which is based on
the concept of generalized non-extensive entropy. Where, a parameter $q$ was introduced to describe the degree of non-extensivity of the system. $q$=1 represents
the classical Boltzmann-Gibbs (BG) statistics; $q$$>$1 leads to an entropy decrease, providing a state of `higher order', whereas for $q$$<$1 the entropy
increases as usual in a closed system, and the system can be considered to evolve towards `disorder'. Based on the Tsallis statistics, a recent
work~\cite{Bertu13} shows that only a small deviation $\delta q$ (lying between (-12$\thicksim$5)\% from the Maxwellian distribution is allowed for the BBN.
However, their treatment did not include the non-extensive impact on the reverse rate, and the lithium problem could not be solved either.

In this work, we have thoroughly calculated the thermonuclear rates for relevant BBN reactions by using the non-extensive $q$-Gaussian distribution.
Here, the non-extensive distribution is only applied to the nucleus, while the photon still obeys the Planck radiation law~\cite{ili07}. As shown by
Torres et al.~\cite{torres}, the leptons and photons well satisfy the classical statistics.
For the first time, the forward and reverse reaction rates have been obtained with the non-extensive distribution coherently. With these non-extensive rates,
the primordial D, $^3$He and $^4$He and $^7$Li abundances are predicted by a BBN code with the up-to-date cosmological parameters and nuclear physics inputs.
Excitingly, it shows that the predicted primordial abundances of D, $^4$He, and $^7$Li agree very well with those observed ones by introducing a non-extensive
parameter in range of 1.063$\leq$$q$$\leq$1.082. Therefore, the pending $^7$Li problem has been solved without introducing any mysterious particles and theory.
Thus, we have discovered that the velocities of nuclei in a hot Big-Bang plasma indeed violate the classical Maxwell-Boltzmann (MB) distribution in a very
small deviation of about 6.3--8.2\%.

It is well-known that thermonuclear rate for a typical $1+2\rightarrow3+4$ reaction is usually calculated by folding the cross section $\sigma(E)_{12}$ with a
MB distribution~\cite{Rolfs88}
\begin{equation}
\label{eq1}
\left\langle\sigma v\right\rangle_{12}=\sqrt{\frac{8}{\pi\mu_{12}(kT)^3}}\int_{0}^{\infty}\sigma(E)_{12}E\mathrm{exp}\left(-\frac{E}{kT}\right)\,dE,
\end{equation}
with $k$ the Boltzmann constant, $\mu_{12}$ the reduced mass of particles $1$ and $2$. In Tsallis statistics, the $q$-Gaussian velocity distribution can be
expressed by~\cite{Silva2,Leubner04}
\begin{equation}
\label{eq2}
f_q(\mathbf{v})=B_q\left(\frac{m}{2\pi kT}\right)^{3/2}\left[1-(q-1)\frac{m\mathbf{v}^2}{2kT}\right]^{\frac{1}{q-1}},
\end{equation}
where $B_q$ denotes the $q$-dependent normalization constant. Thus, the non-extensive reaction rate becomes
\begin{widetext}
\begin{eqnarray}
\label{eq3}
\left\langle\sigma v\right\rangle_{12}= B_q\sqrt{\frac{8}{\pi\mu_{12}}}\times\frac{1}{(kT)^{3/2}}\times\int_{0}^{E_\mathrm{max}}\sigma_{12}(E)E\left[1-(q-1)\frac{E}{kT}\right]^{\frac{1}{q-1}}\,dE,
\end{eqnarray}
\end{widetext}
with $E_\mathrm{max}$=$\frac{kT}{q-1}$ for $q$$>$1, and +$\infty$ for 0$<$$q$$<$1. Here, the $q$$<$0 case is excluded according to the maximum-entropy
principle~\cite{Tsallis88,Tsal09}. Usually, one defines the $1+2\rightarrow3+4$ reaction with positive $Q$ value as the forward reaction, the corresponding
$3+4\rightarrow1+2$ with negative $Q$ value as the reverse one. Under the assumption of classical statistics, the ratio between reverse and forward rates can
be expressed by~\cite{Rolfs88}
\begin{equation}
\label{eq4}
\frac{\left\langle\sigma v\right\rangle_{34}}{\left\langle\sigma v\right\rangle_{12}}=c\times\mathrm{exp}\left(-\frac{Q}{kT}\right),
\end{equation}
with a constant factor defined as $c=\frac{(2J_1+1)(2J_2+1)(1+\delta_{34})}{(2J_3+1)(2J_4+1)(1+\delta_{12})}\left(\frac{\mu_{12}}{\mu_{34}}\right)^{3/2}$.
With Tsallis statistics, however, the reverse rate is expressed as the following equation:
\begin{widetext}
\begin{eqnarray}
\label{eq5}
\left\langle\sigma v\right\rangle_{34}=c\times B_q\sqrt{\frac{8}{\pi\mu_{12}}}\times\frac{1}{(kT)^{3/2}}\times\int_{0}^{E_\mathrm{max}-Q}\sigma_{12}(E)E\left[1-(q-1)\frac{E+Q}{kT}\right]^{\frac{1}{q-1}}\,dE.
\end{eqnarray}
\end{widetext}
The previous work~\cite{Bertu13} determined forward rates using Eq.~\ref{eq3} but then simply determined reverse rates using Eq.~\ref{eq4}. In the
present work we have used a coherent treatment and numerically calculated forward and reverse rates using Eqs.~\ref{eq3} and \ref{eq5}. In addition, the
previous work~\cite{Bertu13} restricted the integral in Eq.~\ref{eq3} to a narrow energy range ($\pm$5$\Delta E_0$). This approximation is actually not
sufficient for large values of $\delta q$. In the present work we have evaluated the integrals in Eqs.~\ref{eq3} and \ref{eq5} without such
restrictions.

As for the typical $1+2\rightarrow\gamma+4$ reaction, the reverse rate becomes
\begin{widetext}
\begin{eqnarray}
\label{eq6}
\frac{\lambda_{\gamma}(4)}{\left\langle\sigma v\right\rangle_{12\to\gamma4}}=\frac{\frac{8\pi\mu_{12}}{h^3}\frac{(2j_1+1)(2j_2+1)}{(2j_4+1)(1+\delta_{12})}\int_{0}^{\infty}\sigma_{12}E \left(e^{\frac{E+Q}{kT}}-1\right)^{-1}dE}
{B_q\sqrt{\frac{8}{\pi\mu_{12}}}{(kT)}^{-3/2}\int_{0}^{E_{max}}\sigma_{12}E\left[1-(q-1)\frac{E}{kT}\right]^{\frac{1}{q-1}}\,dE}.
\end{eqnarray}
\end{widetext}
The photons obey the Planck radiation law in the traditional treatment (e.g., see Refs.~\cite{fcz67,Rolfs88,ili07}), but with an approximation of
$e^{E\gamma/kT}-1\approx e^{E\gamma/kT}$ in calculating the reverse rate. This approximation has been verified recently~\cite{Mathews11} in the BBN temperatures.
Here, we follow the same treatment for photons.

\begin{figure}[tbp]
\begin{center}
\includegraphics[width=8.5cm]{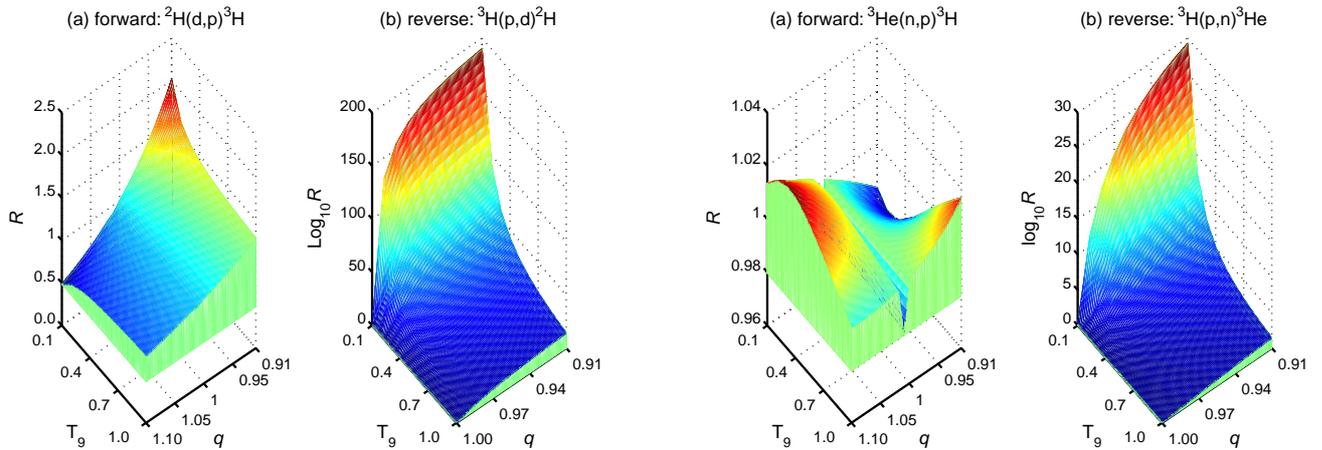}
\vspace{-3mm}
\caption{\label{fig1} (Color online) Ratio between rates calculated using Tsallis and MB distributions for the $^2$H(d,p)$^3$H reaction as functions of
temperature $T_9$ and $q$ values, (a) for forward reaction (in linear scale), and (b) for reverse reaction (in logarithmic scale).}
\end{center}
\end{figure}

\begin{figure}[tbp]
\begin{center}
\includegraphics[width=8.5cm]{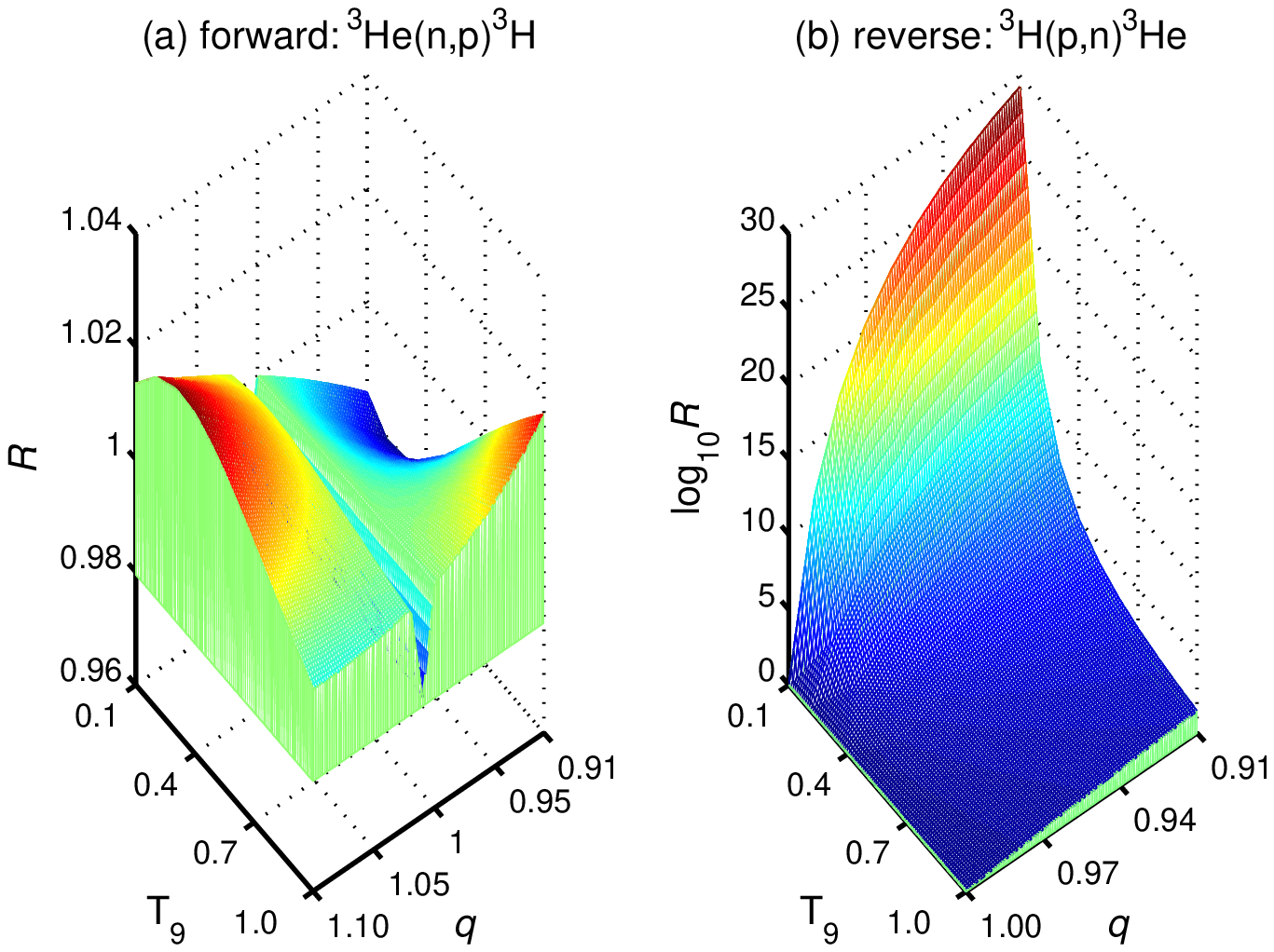}
\vspace{-3mm}
\caption{\label{fig2} (Color online) Results for the $^3$He(n,p)$^3$H reaction, see caption of Fig.~\ref{fig1}.}
\end{center}
\end{figure}

\begin{figure}[tbp]
\begin{center}
\includegraphics[width=8.5cm]{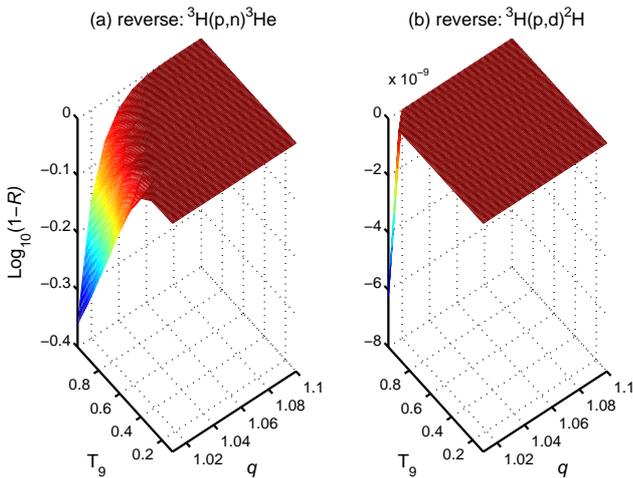}
\vspace{-3mm}
\caption{\label{fig3} (Color online) Results for the reverse rates of $^3$He(n,p)$^3$H and $^2$H(d,p)$^3$H reactions in case of $q$$>$1. See text for details.}
\end{center}
\end{figure}

We show above the impact of $q$ values on the forward and reverse rates of two types of reactions. Here, $^{2}$H(d,p)$^{3}$H is taken as an example of the
charged-particle-induced reaction, and $^3$He(n,p)$^3$H as that of the neutron-induced reaction. Both are among the most important reactions involved in the
BBN. The ratios ($R$) between reaction rates determined with the Tsallis-distribution and MB-distribution are calculated for these two reactions.
Figs.~\ref{fig1} and~\ref{fig2} show the results for forward and reverse rates as functions of temperature and $q$ value. Here, the cross section data for
these two rates are taken from the compilations of Refs.~\cite{Angulo99,Des04}. In the region of 0.1$\leq$$T_9$$\leq$1.0 and 0.91$\leq$$q$$\leq$1.1, the
forward rates calculated with the Tsallis-distribution deviate from the MB rates by relatively modest factors of, at most, 2 and 0.02 for the
$^{2}$H(d,p)$^{3}$H and $^3$He(n,p)$^3$H reactions, respectively. However, the reverse rates for both types of reactions are {\em super}sensitive to deviations
of $q$ from unity. For 0.91$\leq$$q$$\leq$1 (i.e., $q$$<$1), the corresponding Tsallis reverse rates deviate tremendously from the MB rates by about 200 and
30 \textit{orders of magnitude} for $^{2}$H(d,p)$^{3}$H and $^3$He(n,p)$^3$H reactions, respectively. For instance, even with a very small deviation ($q$=0.999),
the Tsallis reverse rate of $^{2}$H(d,p)$^{3}$H is about 10$^{10}$ times larger than the MB reverse rate at 0.2 GK. In order to demonstrate the impact of
$q$$>$1 on the reverse rates of $^3$He(n,p)$^3$H and $^2$H(d,p)$^3$H reactions, the quantities of (1-$R$) are shown in Fig.~\ref{fig3} (in logarithmic scale),
where the top flat plane ($R$=0) represents the reverse rates are negligible small in comparison to the MB rates.

In order to explain qualitatively such {\em super}sensitivity, we define the factor [1-($q$-1)$\frac{E+Q}{kT}$]$^{\frac{1}{q-1}}$ in Eq.~\ref{eq5} as $P_q(E)$.
If $|$1-$q$$|$$\ll$1, $P_q(E)$ can be expressed by the first-order approximation with~\cite{kan98}
\begin{equation}
\label{eq6}
P_q(E) \approx \mathrm{exp}\left[-\frac{E+Q}{kT}+\left(\frac{E+Q}{kT}\right)^2\times \frac{1-q}{2} \right].
\end{equation}
The ratio $R$ defined above for the reverse rates, can then be described approximately by $R>$ exp[($\frac{Q}{kT}$)$^2$$\times$$\frac{1-q}{2}$] for $q$$<$1, and
$R<$ exp[($\frac{Q}{kT}$)$^2$$\times$$\frac{1-q}{2}$] for $q$$>$1. It shows $R$ exponentially depends on the non-extensive parameter $q$, reaction $Q$ value
and temperature. In fact, the sensitivity of $P_q(E)$ (i.e., the tail of distribution) on $q$ results in the huge deviations for the reverse rates compared to
those MB rates. The {\em super}sensitivity of reverse rate on the parameter $q$ has a very important impact: for $q$$<$1, the reverse rates are much larger than
the forward rates, meaning that BBN is increasingly limited in its extent; on the other hand, for $q$$>$1, the reverse rates become negligible compared to the
forward ones, an opposite effect to $q$$<$1.

We have investigated the impact of our new rates on BBN predicted abundances of D, $^3$He and $^4$He and $^7$Li using the code developed in Ref.~\cite{bbn}.
Recent values for cosmological parameters and nuclear physics quantities, such as the baryon-to-photon ratio $\eta$=(6.203$\pm$0.137)$\times$10$^{-10}$~\cite{WMAP9},
and the neutron lifetime $\tau_n=885.7$ s~\cite{nlife}, have been used in our model. The number of light neutrino families $N_{\nu}$=2.9840$\pm$0.0082
determined by CERN LEP experiment~\cite{Nv06} supports the standard model prediction of $N_{\nu}$=3, which is adopted in the present calculation.
The reaction network involves nuclei with $A\leqslant9$ linked by the 34 reactions given in Table~\ref{tab1}. In total, 17 main reactions in the
network~\cite{Smiths93} have been determined using non-extensive statistics, with 11 reactions~\cite{Smiths93} of primary importance and 6 of secondary
importance~\cite{Serpico04} in the primordial light-element nucleosynthesis. The standard MB rates~\cite{wagoner69,cf88,yixu,MFowler,ZHLi,Thomas93} have been
adopted for the other reactions listed. Only those 11 primary important reactions were implemented with the non-extensive statistics in previous
work~\cite{Bertu13}. The present BBN abundances calculated with the usual MB distribution (i.e. $q$=1) agree well with those previous predicted ones
~\cite{coc12,Bertu13,pizz}, which ensures the correctness of the present calculations.

\begin{table}[t]
\caption{\label{tab1} Nuclear reactions involved in the present BBN network. The non-extensive Tsallis distribution is implemented for 17 reactions shown in
bold face. The references for the nuclear physics data adopted for each case are also listed.}
\begin{tabular}{|ll|ll|}
\hline
Reaction & Ref. & Reaction & Ref. \\
\hline
1 n $\to$ p     & \cite{nlife}                                       & 18 \textbf{$^2$H($\alpha,\gamma$)$^6$Li} & \cite{Angulo99,yixu} \\
2 $^3$H$\to^3$He & \cite{Tdecay}                                     & 19$^\ast$ \textbf{$^3$H($\alpha,\gamma$)$^7$Li} & \cite{Des04}  \\
3 $^8$Li$\to$2$^4$He & \cite{Tilley}                                 & 20$^\ast$ \textbf{$^3$He($\alpha,\gamma$)$^7$Be} & \cite{Des04} \\
4 $^6$He$\to^6$Li  & \cite{Tilley}                                   & 21$^\ast$ \textbf{$^2$H(d,n)$^3$He} & \cite{Des04}              \\
5 $^6$Li(n,$\gamma$)$^7$Li & \cite{MFowler}                          & 22$^\ast$ \textbf{$^2$H(d,p)$^3$H} & \cite{Des04}               \\
6 $^2$H(n,$\gamma$)$^3$H   & \cite{wagoner69}                        & 23$^\ast$ \textbf{$^3$H(d,n)$^4$He} & \cite{Des04}              \\
7 $^6$Li(p,$\gamma$)$^7$Be & \cite{yixu}                             & 24$^\ast$ \textbf{$^3$He(d,p)$^4$He} & \cite{Des04}             \\
8 $^6$Li(n,$\alpha$)$^3$H  & \cite{cf88}                             & 25 \textbf{$^7$Be(d,p)2$^4$He} & \cite{Paker72,cf88}            \\
9 $^3$He(n,$\gamma$)$^4$He  & \cite{wagoner69}                       & 26 \textbf{$^7$Li(d,n)2$^4$He} & \cite{cf88}                    \\
10$^\ast$ \textbf{$^1$H(n,$\gamma$)$^2$H}  & \cite{npD}              & 27 $^3$He($^3$He,2p)$^4$He & \cite{cf88}                        \\
11$^\ast$ \textbf{$^3$He(n,p)$^3$H}    & \cite{Des04}                & 28 $^7$Li(n,$\gamma$)$^8$Li & \cite{wagoner69}                  \\
12$^\ast$ \textbf{$^7$Be(n,p)$^7$Li}    & \cite{Des04}               & 29 $^9$Be(p,$\alpha$)$^6$Li & \cite{cf88}                       \\
13$^\ast$ \textbf{$^7$Li(p,$\alpha$)$^4$He} & \cite{Des04}           & 30  2$^4$He(n,$\gamma$)$^9$Be & \cite{cf88}                     \\
14$^\ast$ \textbf{$^2$H(p,$\gamma$)$^3$He}  & \cite{Des04}           & 31 $^8$Li(p,n)2$^4$He & \cite{wagoner69}                        \\
15 \textbf{$^3$H(p,$\gamma$)$^4$He}  & \cite{PTHe4}                  & 32 $^9$Be(p,d)2$^4$He & \cite{cf88}                             \\
16 \textbf{$^6$Li(p,$\alpha$)$^3$He} & \cite{Angulo99,yixu}          & 33 $^8$Li(n,$\gamma$)$^9$Li & \cite{ZHLi}                       \\
17 \textbf{$^7$Be(n,$\alpha$)$^4$He} & \cite{King77}                 & 34 $^9$Li(p,$\alpha$)$^6$He & \cite{Thomas93}                   \\
\hline
\end{tabular}

\footnotesize
$^\ast$: Of primary importance in the primordial BBN~\cite{Smiths93}.
\end{table}

\begin{figure}[tbp]
\begin{center}
\includegraphics[width=8.2cm]{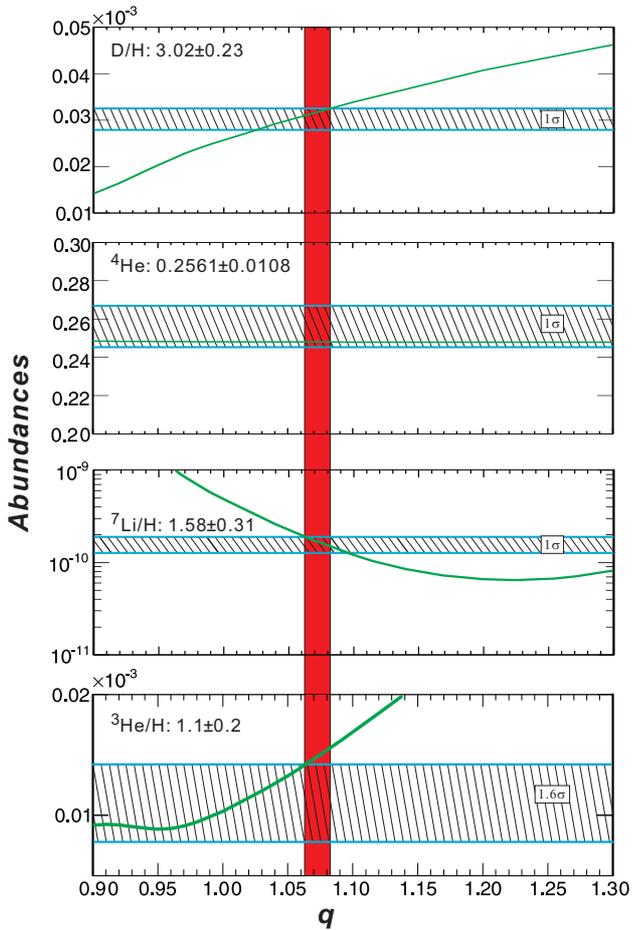}
\vspace{-3mm}
\caption{\label{fig4} (Color online) Predicted primordial abundances v.s. the non-extensive parameter $q$ (in green solid lines). The observed primordial
abundances with 1$\sigma$ uncertainty for D, $^4$He, and $^7$Li are draw together with that with 1.6$\sigma$ for $^3$He. The vertical (red) band constrains the
range of $q$ parameter, i.e., 1.063$\leq$$q$$\leq$1.082, which reconcile the predicted D, $^4$He, and $^7$Li abundances with the observed ones. Note that the
`abundance' of $^4$He exactly refers to its mass fraction.}
\end{center}
\end{figure}

With the new non-extensive rates, we have calculated the D/H, $^3$He/H, $^4$He and $^7$Li abundances as a function of parameter $q$. The results are shown in
Fig.~\ref{fig4}. It shows that the predicted D/H, $^4$He and $^7$Li abundances agree very well with the observations~\cite{oli12,aver10,Li2} in 1$\sigma$
uncertainty in parameter range of 1.063$\leq$$q$$\leq$1.082. The only exception is that $^3$He/H abundance agree with the `observation'~\cite{bania02} in about
1.6$\sigma$ level (see Fig.~\ref{fig4}). Contrary to $^4$He, $^3$He is both produced and destroyed in stars so that the evolution of its abundance as a function
is subject to large uncertainties and has only been observed in our Galaxy. Therefore, as explained before~\cite{coc12} the usually adopted $^3$He/H primordial
abundance observed is not a direct observable, which can not be exactly compared to the BBN prediction.

In fact, the BBN $^7$Li abundance is the sum of final production of $^7$Li and $^7$Be. Since $^7$Be decays to $^7$Li via the electron capture (EC) when the
Universe cooled further. The direct production of these two isotopes are dominated by the radiative capture reactions of $^3$H($\alpha$,$\gamma$)$^7$Li and
$^3$He($\alpha$,$\gamma$)$^7$Be, respectively (see Table~\ref{tab1}). In our model, the baryons are assumed to obey non-extensive distribution, while photons
still follows the classical Planck law. That means the reverse photodisintegration rates for radiative capture reactions keeping the same rate without any
change, but the forward reaction rates are effected greatly by the non-extensive distribution. For the reactions $^3$H($\alpha$,$\gamma$)$^7$Li and
$^3$He($\alpha$,$\gamma$)$^7$Be, the forward rates decrease in case of $q$$>$1 owing to two aspects: the cross sections of both reactions are increased
nearly monotonously with increasing incident energies; the baryon distribution here has a cut-off energy at $\frac{kT}{(q-1)}$ compared to the MB case. In this
circumstance, the creating rates decrease while the destructing rates keep unchanged, resulting in a final considerable reduced production of $^7$Li and $^7$Be.
It is why our predict $^7$Li abundance meets the observations.
Indeed, Clayton et al.~\cite{Clyton75} also proposed an ion distribution in the Sun with a depleted Maxwellian tail (i.e., $q>$1) and `solved' the famous solar
neutrino problem implausibly, although this was later understood to be a consequence of neutrino oscillation.

This work shows that the predicted primordial abundances of D, $^4$He, and $^7$Li agree very well with those observed ones by introducing a non-extensive
parameter $q$. The long-term pending $^7$Li issue is hence solved by the non-extensive statistics without introducing any mysterious particles and
theory. It demonstrates that the velocities of nuclei in a hot Big-Bang plasma indeed violate the classical Maxwell-Boltzmann (MB) distribution by a very small
deviation of about 6.3--8.2\%. The present work also reveals the striking impact that the use of non-extensive statistics might have on calculations of
thermonuclear reaction rates. Indeed, nucleosynthesis in more extreme astrophysical sites such as supernova explosions may be profoundly affected by the
{\em super}sensitivity of endothermic rates on the value of the non-extensive parameter $q$. We encourage extensions of the present study to further interrogate
and test the usual assumptions of classical statistics in even exotic stellar environments (higher-temperature and density astrophysical sites, e.g., nova,
X-ray burst or supernova), which may offer new insight into the nucleosynthesis of heavy elements.

\begin{center}
\textbf{Acknowledgments}
\end{center}
This work was financially supported by the National Natural Science Foundation of China (Nos. 11490562, 11135005, 11321064) and the Major State Basic Research
Development Program of China (2013CB834406). AP was supported by the Spanish MICINN (Nos. AYA2010-15685, EUI2009-04167), by the E.U. FEDER funds as well as by the
ESF EUROCORES Program EuroGENESIS. CB acknowledges support under U.S. DOE Grant DDE- FG02- 08ER41533, U.S. NSF grant PHY-1415656, and the NSF CUSTIPEN grant
DE-FG02-13ER42025. JJ would like to express appreciation to Diego F. Torres (IEEC-CSIC, Barcelona) who made helpful comments on
the manuscript.


\begin{thebibliography}{99}
\bibitem{Li2}
L. Sbordone {\it et al.}, Astron. Astrophys. \textbf{522}, A26 (2010).
\bibitem{Li}
M. Asplund {\it et al.}, Astrophys. J. \textbf{644}, 229 (2006).
\bibitem{Cyburt08}
R.H. Cyburt {\it et al.}, J. Cosmol. Astropart. Phys. \textbf{11}, 012 (2008).
\bibitem{boyd}
R.N. Boyd {\it et al.}, Phys. Rev. D \textbf{82}, 105005 (2010).
\bibitem{kirsebom}
O.S. Kirsebom and B. Davids, Phys. Rev. C \textbf{84}, 058801 (2011).
\bibitem{hamm}
F. Hammache {\it et al.}, Phys. Rev. C \textbf{88}, 062802(R) (2013).
\bibitem{pizz}
R.G. Pizzone {\it et al.}, Astrophys. J. \textbf{786}, 112 (2014).
\bibitem{kusa13}
M. Kusakabe {\it et al.}, Phys. Rev. D \textbf{87}, 085045 (2013).
\bibitem{coc13}
A. Coc {\it et al.}, Phys. Rev. D \textbf{87}, 123530 (2013).
\bibitem{fcz67}
W.A. Fowler, G.R. Caughlan and B.A. Zimmerman, Ann. Rev. Astron. Astrophys. \textbf{5}, 525 (1967).
\bibitem{Wagoner67}
R.V. Wagoner, W.A. Fowler and F. Hoyle, Astrophys. J. \textbf{148}, 3 (1967).
\bibitem{wagoner69}
R.V. Wagoner, Astrophys. J. Suppl. \textbf{18}, 247 (1969).
\bibitem{cf88}
G.R. Caughlan and W.A. Fowler, At. Data Nucl. Data Tables \textbf{40}, 283 (1988).
\bibitem{Smiths93}
M.S. Smith, L.H. Kawano and R.A. Malaney, Astrophys. J. Suppl. \textbf{85}, 219 (1993).
\bibitem{Angulo99}
C. Angulo {\it et al.}, Nucl. Phys. \textbf{A656}, 3 (1999).
\bibitem{Des04}
P. Descouvemont {\it et al.}, At. Data Nucl. Data Tables \textbf{88}, 203 (2004).
\bibitem{Serpico04}
P.D. Serpico {\it et al.}, J. Cosmol. Astropart. Phys. \textbf{12}, 010 (2004).
\bibitem{yixu}
Y. Xu {\it et al.}, Nucl. Phys. \textbf{A918}, 61 (2013).
\bibitem{mandl08}
F. Mandl, Statistical Physics (2nd Edition), John Wiley \& Sons, 2008.
\bibitem{Rolfs88}
C.E. Rolfs and W.S. Rodney, Cauldrons in the Cosmos, University of Chicago Press, 1988.
\bibitem{miller55}
R.C. Miller and P. Kusch, Phys. Rev. \textbf{99}, 1314 (1955).
\bibitem{tar02}
A. Taruya and M. Sakagami, Physica A \textbf{307}, 185 (2002).
\bibitem{Tsallis88}
C. Tsallis, J. Stat. Phys. \textbf{52}, 479 (1988).
\bibitem{Tsal09}
C. Tsallis, Introduction to Nonextensive Statistical Mechanics, Springer Verlag, 2009.
\bibitem{Bertu13}
C.A. Bertulani {\it et al.}, Astrophys. J. \textbf{767}, 67 (2013).
\bibitem{ili07}
C. Iliadis, Nuclear Physics of Stars, WILEY-VCH, Weinheim, 2007.
\bibitem{torres}
D.F. Torres {\it et al.}, Phys. Rev. Lett. \textbf{79}, 1588 (1997), Erratum: ibid, \textbf{80}, 3889 (1998).
\bibitem{Silva2}
R. Silva Jr. {\it et al.}, Phys. Lett. A \textbf{249}, 401 (1998).
\bibitem{Leubner04}
M.P. Leubner, Astrophys. J. \textbf{604}, 469 (2004).
\bibitem{Mathews11}
G.J. Mathews {\it et al.}, Astrophys. J. \textbf{727}, 10 (2011).
\bibitem{kan98}
G. Kaniadakis {\it et al.}, Physica A \textbf{261}, 359 (1998).
\bibitem{bbn}
S.Q. Hou {\it et al.}, Chin. Phys. Lett. \textbf{27}, 082601 (2010).
\bibitem{WMAP9}
G. Hinshaw {\it et al.}, Astrophys. J. Suppl. \textbf{208}, 19 (2013).
\bibitem{nlife}
Particle Data Group, J. Phys. G: Nucl. Part. Phys. \textbf{33}, 1 (2006); Phys. Lett. \textbf{B667}, 1 (2008); J. Phys. G: Nucl. Part. Phys. \textbf{37}, 075021 (2010).
\bibitem{Nv06}
LEP Collaboration, Phys. Rep. \textbf{427}, 257 (2006).
\bibitem{Tdecay}
L.L. Lucas and M.P. Unterweger, J. Res. Natl. Inst. Stand. Tech. \textbf{105}, 541 (2000).
\bibitem{Tilley}
D.R. Tilley {\it et al.}, Nucl. Phys. \textbf{A708}, 3 (2002); ibid \textbf{A745}, 155 (2004).
\bibitem{MFowler}
R.A. Malaney and W.A. Fowler, Astrophys. J. \textbf{345}, L5 (1989).
\bibitem{npD}
K.Y. Hara {\it et al.}, Phys. Rev. D \textbf{68}, 072001, (2003).
\bibitem{PTHe4}
S.B. Dubovichenko, Rus. Phys. J. \textbf{52}, 294 (2009).
\bibitem{King77}
C.H. King {\it et al.}, Phys. Rev. C \textbf{16}, 1712 (1977).
\bibitem{Paker72}
P.D. Parker, Astrophys. J. \textbf{175}, 261 (1972).
\bibitem{ZHLi}
Z.H. Li {\it et al.}, Phys. Rev. C \textbf{71}, 052801(R) (2005).
\bibitem{Thomas93}
T. Thomas {\it et al.}, Astrophys. J. \textbf{406}, 509 (1993).
\bibitem{coc12}
A. Coc {\it et al.}, Astrophys. J. \textbf{744}, 158 (2012).
\bibitem{aver10}
E. Aver {\it et al.}, J. Cosmol. Astropart. Phys. \textbf{5}, 003 (2010).
\bibitem{oli12}
K.A. Olive {\it et al.}, Mon. Not. R. Astron. Soc. \textbf{426}, 1427 (2012).
\bibitem{bania02}
T.M. Bania, R.T. Rood and D.S. Balser, Nature \textbf{415}, 54 (2002).
\bibitem{Clyton75}
D.D. Clayton {\it et al.}, Astrophys. J. \textbf{199}, 494 (1975).

\end{thebibliography}
\end{document}